\documentclass[a4paper,12pt]{article}
\usepackage{graphicx}

\begin{document}

\title{Forty years of acting electron-positron colliders}
\author{V. N. Baier\\
Budker Institute of Nuclear Physics,\\
Novosibirsk, 630090, Russia}

\maketitle
\begin{abstract}
Around forty years passed from the beginning of operation of the
first electron-positron colliding beam facility VEPP-2 in Institute
of Nuclear Physics (INP), Novosibirsk. Here I described development
of electron-positron colliding beam project in INP, as well as
advance of similar projects of the first generation at LAL, Orsay
and at LNF, Frascati.

\end{abstract}

\section{Introduction}

In the mid of 50's of last century electron accelerators of higher
and higher energy were built, while their usefulness was limited by
the fact the energy available for physics study, e.g. for analysis
of interaction at small distances or for creating of new particles,
is that measured in the center-of-mass system (c.m.s.) of target and
incident electron. In the relativistic limit the energy in c.m.s. is
$\varepsilon_c=\sqrt{2\varepsilon_l mc^2}/2$, where $\varepsilon_l$
is the electron energy in the laboratory system, $m$ is the mass of
target. In electron-electron interaction for $\varepsilon_l$=6~GeV
(the highest electron energy planned at that time) one has only
$\varepsilon_c$=39~ MeV. Because of this reason a new idea appeared:
to create storage rings where two electron beams (or two proton
beams) traveling in opposite directions could collide. The colliding
beams idea was proposed nearly simultaneously in 1956 \cite{K},
\cite{N} and actively discussed at 1956 CERN Accelerator Symposium.
After several laboratories started serious $R@D$ and design
activity. SLAC proposal to study limits of quantum electrodynamics
was issued in May 1958 \cite{NBRP}.

After completing my post-graduated courses at Moscow Lebedev
Institute at January 1959 I joined  newly organized Institute of
Nuclear Physics (INP) of Siberian Branch of Academy of Science of
USSR\footnote{Decision of Soviet Government about creation of the
new large Scientific Center near Novosibirsk devoted to development
of Siberia was issued May 18, 1957. Formal decision about creation
Nuclear Physics Institute was issued February 21, 1958.}. Professor
G.I.Budker was appointed director of the institute. This was
40-years ambitious physicist working on fairly nonstandard problems
of plasma and accelerator physics. The institute was formed on the
basis of small Budker's laboratory "New  acceleration methods" at
Nuclear Energy Institute (now Kurchatov Institute) in Moscow. At the
end of 1958 scientific staff of Institute was consisted of around 30
persons most of them were graduated students. The very new idea of
electron-electron colliding beams, for realization of which many
original development were needed, attracted Budkers's attention and
creation of electron-electron colliding beam facility became one of
main purposes of newly created institute.

In 1959 the colliding beams team worked on technical design of
facility including some specific issues such as fast injection
kicker magnets, ultra-high vacuum system, detector of scattered
electrons, etc. At the same time the program of physics research was
under development. Budker seek for support of colliding beams
project and invited known scientists to discuss the research
program. In October 1959 I.Ya.Pomeranchuk visited institute. In
Budker's office it was quite long discussion of electron-electron
colliding beams facility where study of quantum electrodynamics at
small distances was mixed with technical details of installation,
which was at very beginning of construction. Pomeranchuk was not in
raptures concerning the discussion and no support to the project was
expressed. Besides other participants of the meeting did not express
any enthusiasm. After Pomeranchuk left institute, Budker came to my
office, he was complaining that institute research program is not
enough impressive and one has to think how improve it. I replied
that the program becomes immeasurably more rich, if one creates
electron-positron colliding beams. "You are mad!" said Budker and
left. But several minutes later he arrived back: "Tell me once
more!"  In very intense discussion which followed we expressed many
times the pros and cons realization of electron-positron colliding
beams. In the end, late in the evening, Budker demanded that I had
to lay aside all my business and concentrate on realization of new
proposal. It was October 28, 1959.

The next day active work began. We considered the electron-positron
colliding beams installation step by step from morning to late
evening.  Within a week the very rough draft of facility which later
became VEPP-2 was prepared. The maximal energy per beam (700~MeV)
was selected to produce K-mesons. Special attention was devoted to
positron production by electrons in tungsten converter. The first
design drawing of electron-positron facility were made in December
1959. It should be noted that at that time all electron-electron
colliding beam projects were on the very preliminary stage.

 The
research program which was formulated at that time was \cite{B}:
\begin{enumerate}
\item Study of elastic electron-positron scattering at large angle
to test QED at small distances (similar to a goal of
electron-electron colliding beams).
\item Study of annihilation electron-positron pair into two photons.
This is an additional test of QED.
\item Study of conversion of $e^+e^-$ pair into pair of $\mu^+$ and
$\mu^-$ mesons. This is a test if the $\mu$ meson is pointlike and
if so this is an additional channel to test QED.
\item
Study of conversion of $e^+e^-$ pair into pair of $\pi^+$ and
$\pi^-$ mesons to study the electromagnetic form-factor of pion at
time-like momentum transfer.
\item
Study of conversion of $e^+e^-$ pair into pair of $K^+$ and $K^-$
mesons to study the electromagnetic form-factor of kaon at time-like
momentum transfer.
\end{enumerate}

It is reasonable to describe the scene in elementary physics
research at that time. A few electron synchrotrons with energy
slightly higher than 1~GeV were brought into operation in middle of
50's, so the experience of work with such accelerators was quite
limited. The table of elementary particles contained leptons:
electron, positron, neutrino, antineutrino, $\mu^{\pm}$ meson; and
strongly-interacting particles: proton, neutron, a few types of
hyperons, as well as $\pi^{\pm}$ and $\pi^0$ mesons and different
types of K mesons.

It is evident, that electron-positron colliding beams could give
very new opportunities not only for test of QED at small distances
but also for study of electromagnetic properties of mesons
participating in strong interaction. At that time there was no
information at all about these properties. Even the cross section of
such simple process as transformation of electron-positron pair onto
pair of $\mu^+$ and $\mu^-$ mesons was calculated by Berestetsky and
Pomeranchuk only in 1955 \cite{BP}.

How to realize electron-positron facility was the question. First of
all it was necessary to develop positron production system.
Positrons (antiparticles) were observed in cosmic rays and in
reaction at accelerators. However considerable amount of positrons
was never produced. In the 30's-40's the theory of electron-photon
showers was developed. This was a basement. But for
electron-positron facility one has to produce {\it beam} of
positrons. The elaborated scheme of positron production is used now
everywhere, naturally with many perfection: the electron beam with
energy a few hundreds MeV is directed to heavy metal (e.g. tungsten)
target with thickness 1-2 radiation length. In the target electron
radiate a photon in collision with a nucleus, then this photon
interacts with another nucleus and creates electron-positron pair.
Created positrons should be collected and accelerated and then
injected to storage ring. We started calculation of conversion of
electrons into positrons nearly from the very beginning of project.
The good peace of this work was done by Synakh \cite{Sy} (at that
time my post-graduated student). For example, the calculated
conversion coefficient $\mu$ of electrons with energy
$\varepsilon_0=500$~MeV into positrons with energy $\varepsilon =
250$~MeV into the energy interval $\Delta \varepsilon/\varepsilon
\sim 5$\% is $\mu \simeq 1/400$ for the converter with thickness
around 1 radiation length. The created positrons are moving mostly
ahead in the direction of electron momentum. In the mentioned
example the angle of positron cone is $\sim 4^{\circ}$. So, other
things being equal, for production of positron beam of some
intensity one needs the electron beam which is thousand more
intense. If one accumulates positrons by small bunches, then the
storage time will be thousand times longer.

A few graduated student were recruited to the team which started the
development of electron-positron facility. They began more detail
study of the project including electron injector, ejection from it
using the fast kicker magnets, channels and conversion of electrons
into positrons in one of channels, injection into sole storage ring
where electrons and positrons are moving in opposite directions,
ultra-high vacuum system in storage ring and stability of orbits in
it.

At that time (end of 1959) the electron-electron colliding beam
projects were only at very preliminary stage of development and some
people doubted that the colliding beam technology could be used in
high-energy physics research. In this situation the proposal to
build the electron-positron facility with essentially higher
requirements for intensity and quality of beams was coldly received
by many famous members of Soviet Academy of Science. Only support of
I.V.Kurchatov at that time very influential director of Nuclear
Energy Institute permitted to start development of the project.
Another factor which lowered level of opposition to the project was
transfer of the all team from Moscow to the wild East: to
Novosibirsk, where building of edifices of Nuclear Physics Institute
began in 1959.

In 1960 when the staff was still very small all members of the team
were working on various topics of both electron-electron and
electron-positron installations. Alexander Skrinsky was appointed as
a head of laboratory. I have found not long ago the internal report
of INP-1960 with a title "Motion of particles in an accelerator with
racetrack" by V.N.Baier, V.S.Synakh and I.B.Khriplovich devoted to
study of particle dynamics in VEPP-2.

In 1961 the main part of staff (including Budker, Skrinsky and
myself) was moved from Moscow to the new Academic town
(Academgorodok) in 30 km south from Novosibirsk, where the first
building of INP was constructed. Evidently, transfer of the team
with equipment was an obstacle in the way of project development.
However, it was quite good financial support of the new Institute in
Novosibirsk and this permitted to create quite effective Institute's
workshop for manufacturing of parts of the facility, and to order
most big parts in Novosibirsk industry from the one side, and to
recruit many graduated students from Novosibirsk universities from
the other side.

In the beginning of 1961 our library received {\it Il Nouvo Cimento}
with announcement about Frascati storage ring \cite{BCT}. This
showed that we were not alone in the field\footnote{ Contacts with
foreign laboratories were so limited at that time, that the
scientific information one drew only from journals.}. But from point
of view of our experience at that time it was evident (we are worked
hard at injection system and creation of positron beam) that at best
the very limited amount of electrons and (or) positrons could be
stored in storage ring without direct particle injection.

In 1962 some parts of VEPP-2 were ready and tested. The storage ring
VEP-1 manufactured at big Novosibirsk factory was first moved in
1961 to the Kurchatov Institute in Moscow, then in summer of 1962 it
was disassembled together with synchrotron B-2S which was its
injector and moved to INP in Novosibirsk where both was assembled
fast and immediately the test operations began.

In the year 1963 it was permitted the complete legalization of
activity of INP\footnote{Because of pathological secrecy adopted at
that time in USSR, all activity in Kurchatov Institute was
considered as "for restricted use only" and the special permission
for publication in open journals or proceedings was necessary for
each article.}. At the International Conference on high-energy
accelerators in Dubna in August 1963 the collider activity of INP
was presented for the first time \cite{INP1}. The photo of VEPP-2
assembly process was contained in the report along with other
information. The main contributors are the authors of the
corresponding parts.

Before the Conference we got information that the Frascati storage
ring AdA moved from Frascati to Orsay, where there was the
high-energy linac which was used as an injector. The new interesting
effect was observed: the loss of particles in storage ring due to
electron-electron scattering inside the bunch \cite{BCG}. Under some
conditions the lifetime of a beam in a storage ring is determined
just by this event which is now called the Touschek effect.

We have learned at the Conference that two more teams started the
work in the same direction. These were the projects of Orsay
electron-positron storage ring ACO with energy up to 450 MeV
 at Laboratoire  de l'Acc$\acute{\rm e}$l$\acute{\rm
e}$rateur Lin$\acute{\rm e}$aire (LAL) in Orsay, France \cite{SR}
and electron-positron storage ring ADONE with energy up to 1.5 GeV
at Laboratori Nazionali di Frascati  (LNF) in Frascati, Italy
\cite{AA}. Somewhat later we received description of ACO in detail
\cite{ADS}.

This indicated that creation of electron-positron colliders became
very respectable and perspective direction of development of
high-energy accelerators. Some kind of race emerged for the physics
meaningful results at these colliders.

The study of electron-electron scattering at VEP-1 storage ring
began in 1964. The first small angle scattering event was registered
on May 19.

A very important characteristics of colliding beam facility is {\it
the luminosity $L$}. The number of events per unit time (usually 1
sec) $N$ of some process with the total cross section $\sigma$ is
$N=L\sigma$. The luminosity is proportional to product of current in
the beams and inverse proportional to the transverse section of beam
$S$. To obtain acceptable luminosity one has to work with enough
high currents and small size beams.  A very high vacuum and damping
of beam instabilities are the necessary conditions to have
reasonable small beam size.

A new type of sophisticated detectors had to be created. Ben Sidorov
was in charge of this direction in INP.

The data taking at VEPP-2 installation began in 1966 \cite{ABN}. The
results will be discussed below.

 \section{Physics with electron-positron colliding beams}
 \subsection{Radiative corrections}

 On the first stage of the electron-positron project development one
 of main goals was the check of the applicability of quantum
electrodynamics at small distances. The qualitative estimate of
measured distance\footnote{Below the system of units where
$\hbar=c=1$ is used.} is $\lambda \sim 1/q=(\hbar/q)$, where $q$ is
the momentum transfer. For $q \sim 1~$GeV the measured distance is
$\lambda\sim 0.2~$fm \newline(1 fm=$10^{-13}$ cm is a typical hadron
scale).

The cross sections of electron-electron and electron-positron
scattering in Born approximation (order $\alpha^2$) was calculated
by M{\o}ller and Bhabha in 30's. The fantastic development of
Quantum Electrodynamics (QED) in 40's permitted consideration of
higher order corrections (the series with respect to powers
$\alpha=e^2$) which are called radiative corrections(RC). In the
late 50's these corrections to the mentioned cross sections were the
topic of QED textbook (e.g. \cite{JR, AB}).  At high energy
$\varepsilon \gg m$ the actual parameter of decomposition is
$(\alpha/\pi)\ln (\varepsilon/m)$. Calculation of RC includes
obligatory (because of infrared divergence) contribution from
radiation of real photons and because of this depends on the
particular experimental set-up.  In the specific conditions of
binary ($2\rightarrow2$) reactions on colliding beams the photon
emission from one of initial particles causes non-collinearity of
the final particle momenta $\Delta\vartheta$. Since for elastic
cross section the events with minimal $\Delta\vartheta$ are
selected, this imposes substantial limitation on energy
$\Delta\varepsilon$ radiated from the initial particles. Accuracy of
measurement of final particles energy in the 1st generation detector
was quite poor. This means that hard photon emission from final
particles are allowed and this complicates calculation. The cross
section with the radiative corrections $\delta_R$ taken into account
are usually written in the form: $d\sigma=d\sigma_0(1-\delta_R)$,
where $d\sigma_0$ is the M{\o}ller (or Bhabha) cross section.  The
main term of RC of the lowest order $\propto \alpha^3$ (so-called
"double-logarithm term" containing the product of two large
logarithm: logarithm of energy and logarithm of ratio $\Delta
\varepsilon/\varepsilon$, which arises from sum of contributions of
soft virtual and real photons) is $\delta_R = (8\alpha/\pi)
\ln(\varepsilon/m)\ln \varepsilon/\Delta\varepsilon)$, where
$\Delta\varepsilon$ is the total energy of emitted quanta. The
different aspects of radiative effects in electron-electron
(positron) collisions were analyzed with Sam Kheifets, Victor Fadin
and Valery Khoze. The complete expressions for $\delta_R$ in $e-e$
scattering are given in \cite{BKh, BFK} and in $e^+-e^-$ scattering
in \cite{SFK}.

For typical experimental conditions at $\varepsilon \sim 1$~GeV one
has the radiative correction $\delta_R \sim 10\%$ and evidently it
should be taken into account in comparison of theory and data.

At high energies and for $\Delta \varepsilon/\varepsilon \ll 1$ the
soft-photon corrections dominate, e.g. for $\varepsilon=7$~GeV and
$\Delta \varepsilon/\varepsilon=10^{-2}$ one has $\delta_R=0.75$,
and one can't be restricted to the lowest order of perturbation
theory. So the general analysis of RC in the all orders of
perturbation theory is of significant interest. In the 50's it was
fashionable to study structure of QED as a whole. The method of
calculation of the cross sections in high energy QED proposed by
Abrikosov \cite{A} was used. Within framework of this method, only
those contributions are retained which contain the maximum power of
the large logarithms. For test of QED at small distances only
processes with large momentum transfer are of interest. It's
remarkable that in this case in any order of the perturbation theory
only contributions of diagrams with one photon exchange between
charged particle lines survives, while all other contributions
cancel each other, and in double-logarithm approximation the
scattering cross section after inclusion of soft photon emission
acquires the form
$d\sigma(\vartheta)=d\sigma_0(\vartheta)e^{-\delta_R}$ \cite{BKh1,
BKh2}. It was shown that in double-logarithm approximation  the
cross sections of all processes of scattering and pair creation for
large momentum transfers have soft-photon nature (see \cite{BKh1,
BKh2, G}). As it was mentioned, only diagrams with one photon
exchange between charged particles contribute. An interesting
application is the behavior of $e^-+e^+ \rightarrow \mu^-+\mu^+$
cross section  near threshold. In the case when there is no
limitation on photon emission $\delta_R \rightarrow
\delta_{\mu}=(4\alpha/\pi)\ln(\varepsilon/(\varepsilon-\mu))\ln(\varepsilon/m)$,
where $\mu$ is the muon mass. For nonrelativistic muons one has
\cite{BKh2}
\begin{equation}
\sigma(\vartheta) =
\sigma_0(\vartheta)\left(\frac{q}{\sqrt{2\varepsilon
\mu}}\right)^{\frac{8\alpha}{\pi}\ln\frac{\varepsilon}{m}}|\psi(0)|^2,
\quad \sigma_0(\vartheta)=\frac{\alpha^2q}{16\varepsilon^3}\left[1+
\frac{\mu^2}{\varepsilon^2}+\frac{q^2}{\varepsilon^2}\cos^2\vartheta
\right], \label{1}
\end{equation}
where $q$ is the momentum of final muon and $\vartheta$ is the angle
between momenta of initial electron and final $\mu^-$, factor
$|\psi(0)|^2$ takes into account the Coulomb interaction between
final particles found by Sakharov \cite{Sak}:
$|\psi(0)|^2=(2\pi\alpha/v)/(1-e^{-2\pi\alpha/v})$, here $v$ is the
muon velocity. In the region where the Coulomb interaction is
insignificant ($2\pi\alpha/v \gg 1$) one obtains $\sigma \propto
q^{1.1}$ in place of $\sigma \propto q$; and for $\tau$-lepton
$\sigma \propto q^{1.15}$.

The test of the applicability of QED at small distances by
electron-electron scattering was performed at VEP-1 storage ring in
Novosibirsk \cite{Bud} and at Princeton-Stanford storage ring
\cite{STG}, \cite{STR}. The data \cite{STG}, \cite{STR} have been
compared with the M{\o}ller formula modified by a form factor
 $f(q^2)$ (in vertex $\gamma_{\mu} \rightarrow \gamma_{\mu}f(q^2)$ and
$1/q^2\rightarrow f(q^2)/q^2$) and usually a simple parametrization
is used\footnote{For more general modification see \cite{B1}}
$f^2(q^2)=1/(1\pm q^2/\Lambda_{\pm}^2)$  and with radiative
correction $\delta_R$ , calculated by Tsai  \cite{T0,T}, taken into
account. The limit $\Lambda=\infty$ corresponds to interaction of
point particles. The results of experiment \cite{STR} are $\Lambda_-
> 4.4$~GeV and $\Lambda_+ > 2.7$~GeV (95\% confidence). The limit
$\Lambda_- > 4.4$~GeV means that QED was checked for distance $l
\geq$ 0.05~fm. The distance $l$ is much shorter than characteristic
hadronic dimension.

Similar limitations where obtained for different processes at
electron-positron colliders of the first generation (95\%
confidence):
\begin{enumerate}
\item Electron-positron elastic scattering measured in Orsay
(ACO collider) with beam energy $\varepsilon=510$~MeV \cite{Aug}
with RC $\delta_R \sim~7\%$ results $\Lambda_- > 3.8$~GeV and
$\Lambda_+ > 2.8$~GeV and measured in Frascati (Adone collider) with
beam energy $\varepsilon=700-1200$~MeV \cite{Ber} gives
$\overline{\Lambda_{\pm}} > 6$~GeV.
\item Two-photon annihilation of electron-positron pair
measured on VEPP-2 collider in Novosibirsk with beam energy
$\varepsilon=500$~MeV \cite{Bal1} results $\overline{\Lambda_{\pm}}
> 1.3$~GeV and measured in Frascati (Adone collider) with beam
energy $\varepsilon=700-1200$~MeV \cite{Ber} gives $\Lambda_- >
2.0$~GeV and $\Lambda_+ > 2.6$~GeV. In this reaction both the
modifications of vertex and of electron propagator in a consistent
(gauge invariant) way were introduced.
\item Study of conversion of $e^+e^-$ pair into pair of $\mu^+$ and
$\mu^-$ mesons on VEPP-2 collider in Novosibirsk with beam energy
$\varepsilon=500$~MeV \cite{Bal2} gives $\overline{\Lambda_{\pm}} >
3.1$~GeV (95\% confidence) and measured in Frascati (Adone collider)
with beam energy $\varepsilon=700-1200$~MeV \cite{Ber} gives
$\Lambda_- > 5$~GeV.
\end{enumerate}

The tests of QED at small distances were continued at next
generations of electron-positron colliders, which were built later
in Novosibirsk, Stanford, Cornell, Orsay, Frascati, Tsukuba, Geneva.
The described above limitations were substantially improved (by two
orders of magnitude) due to higher energy, larger circulating
current and better detectors. For example, the two(three)-photon
annihilation of electron-positron pair\footnote{To elastic $e^+e^-$
scattering and to the process  $e^+e^-\rightarrow\mu^+\mu^-$ there
is an additional (and quite significant for used energy)
contribution from $Z$ boson. So these processes can't be used for
pure QED test.} measured on LEP collider in CERN (Geneva) with beam
energy $\varepsilon=45-101$~GeV \cite{Zh} gives $\Lambda_-
> 258$~GeV and $\Lambda_+ > 415$~GeV. The last limit
means that QED is checked for distance $l \geq 5\cdot 10^{-17}$~cm.

\subsection{Inelastic processes}

At low energy $\varepsilon \sim m$ the electromagnetic processes are
sorted usually over powers of fine-structure constant
$\alpha=e^2=1/137$  in frame of perturbation theory. The cross
sections of simplest two-particle processes:
electron(positron)-electron scattering, photon-electron scattering,
annihilation of electron-positron pair into two photon or pair of
charged particles are of order $\alpha^2/m^2 =r_0^2 \sim 10^{-25}
{\rm cm}^2$. In many-particle processes each additional particle
adds factor $\alpha$ to the cross sections. Such processes were
considered only in the form of RC, as it was discussed above. At
high energy $\varepsilon \gg m$ the situation changes; the magnitude
of the cross sections is determined mainly by the dependence on
energy. Understanding of importance of such classification arose
along with development colliding beam program. The processes,
diagram of which contains two blocks (each of which is attached to
charged particle line) connected with photon(photons) line, have
nondecreasing as a function of energy total cross section. Besides
the power constancy in some cases there are the logarithmic growth
with energy.\footnote{ The only process which is fallen out this
scheme is the elastic electron(positron)-electron scattering where
the total cross sections diverges at any energy.} The important
example is the process of soft $n$-photon radiation in
electron(positron)-electron scattering which was studied by Victor
Galitsky and I \cite{BG}. The emission of classical ($\omega \ll
\varepsilon$) photons occurs in an independent way so that the cross
section of process with the emission of $n$ photons may be
represented as, see e.g. \cite{YFS}:
$d\sigma_n=d\sigma_0\prod_{i=1}^n dW(k_i)/n!$, where $\sigma_0$ is
the cross section of elastic process,
$dW(k_i)=|j|^2d^3k_i/2\omega_i$ here $j_{\mu}$ is the "classical"
current, for electron scattering off Coulomb center
$j_{\mu}=i\sqrt{2}e(p_{\mu}/(kp)-p'_{\mu}/(kp'))/(2\pi)$, where
$p_{\mu} (p'_{\mu})$ is the initial (final) electron momentum,
generally each line of charged particle in the process diagram
contributes to the current the combination $\pm p_{\mu}/(kp)$. For
bremsstrahlung $d\sigma_1=d\sigma_0 dW(k)$, where $d\sigma_0$ is the
Rutherford formula. Integrating $dW(k)$ over photon emission angles
$\Omega$ one obtains
\begin{equation}
dI(\omega, x)=\int dW(k)=
\frac{2\alpha}{\pi}\frac{d\omega}{\omega}\Phi(x), \quad \Phi(x)=
\frac{2x^2+1}{x\sqrt{1+x^2}}\ln(x+\sqrt{1+x^2})-1, \label{11}
\end{equation}
where $4m^2x^2=-(p-p')^2=4{\bf p}^2\sin^2(\vartheta/2),~\vartheta$
is the electron scattering angle. In the limiting cases one has: $x
\ll 1,~\Phi(x)=4x^2/3$ and $x \gg 1,~\Phi(x)=\ln4x^2 - 1$. The
universal function $\Phi(x)$ defines the probability dependence on
the momentum transfer in soft photon radiation.

To find the integral cross section one has to integrate $d\sigma_0
dI(\omega, x)$ over the momentum transfer $x$. Taking into account
that the Rutherford cross section $d\sigma_0=(\pi Z^2
\alpha^2/m^2)dx^2/x^4$ it is clear that the main contribution gives
region $x \ll 1$. The minimal value of $x$ is attained when all the
momenta are collinear: $4m^2x^2_{min}=\omega^2m^4/4\varepsilon^4$.
Within logarithmic accuracy one can put $x^2_{max}=1$. Substituting
the functions in $d\sigma_0 dI(\omega, x)$ for $x \ll 1$ and
performing integration one find the spectrum of bremsstrahlung of
the photon with energy $\omega$:\newline
$\sigma_1=(16/3)(Z^2\alpha^3/m^2)(d\omega/\omega)\ln(4\varepsilon^2/m^2\omega)$.
The region $x \gg 1$ does not contribute because of fast decreasing
of the Rutherford cross section with $x$ increase.

The result of similar analysis for bremsstrahlung at
electron-electron(positron) scattering ($Z=1$) differs from this
expression only by logarithm argument:$4\varepsilon^2/m^2\omega
\rightarrow 8\varepsilon^3/m^3\omega$ and radiation takes place in
the direction of motion of both colliding particles.

In the case of many photons radiation the integration over photon
emission angles can be performed independently, so that
$d\sigma_n=d\sigma_0\prod_{i=1}^n dI(\omega_i, x)/n!$. Of course,for
electron-electron(positron) scattering in c.m.s. one has to use the
corresponding classical current. However the only region of small
momentum transfer contributes and final results is expressed in
terms of the function $\Phi(x)$. Since at $x \ll 1,~\Phi(x) \propto
x^2$ starting from $n \geq 2$ there is no divergence at $x=0$ and
with high accuracy one can put $x_{min}=0$. So, 1) the cross section
$d\sigma_n$ does not contain large logarithm; 2) the value
$d\sigma_n$ can be calculated within power accuracy (discarded terms
$\sim m^2/\varepsilon^2$); 3) the main contribution gives region $x
\sim 1$; 4) at $x \gg 1$ the cross section looks like $\ln^n x
dx^2/x^4$ and because convergence of the integral one can put
$x_{max}=\infty$ within accuracy $\sim m^2/\varepsilon^2$. Thus, the
cross section of $n$ photon radiation at electron-electron(positron)
collision is
\begin{equation}
d\sigma^e_n=2\pi\frac{\alpha^2}{m^2}
\left(\frac{4\alpha}{\pi}\right)^n\frac{\nu(n)}{n!}
\prod_{i=1}^n\frac{d\omega_i}{\omega_i}, ~\nu(n)=\int_0^{\infty}
\Phi^n(x)\frac{dx}{x^3}, \label{12}
\end{equation}
here
$\nu(2)=5/4+7\zeta(3)/8,~\nu(3)=3[8\zeta(3)-1]/5,~\zeta(3)=1.202..$
is Riemann's zeta function. The simple combinatory analysis shows
that when $m$ photons are emitted in the direction of one particle
and $n-m$ photons are emitted in the direction of another particle
the corresponding cross section is $d\sigma^e_n(m,
n-m)=C^n_md\sigma^e_n/2^n$. For the double bremsstrahlung in the
case when photons are emitted in opposite directions
$d\sigma^e_2(1,1)=d\sigma^e_2/2$ and in the case when both photons
are emitted in the one direction
$d\sigma^e_2(2,0)=d\sigma^e_2(0,2)=d\sigma^e_2/4$.

So considered cross section grows logarithmically with energy
increase at $n=1$, while at $n \geq 2$ it doesn't dependent on
energy.

\subsubsection{Single bremsstrahlung in electron-(electron)positron collision}

This simplest inelastic process is represented by 8 Feynman diagrams
and the differential cross section is very cumbersome. In
high-energy region which is of main interest, one can decompose
cross section over powers of $m/\varepsilon$. Moreover the
calculation simplified essentially if one integrates contributions
of the radiation block of diagram in tensor form taking into account
invariance properties of QED \cite{BG1}. In the center of mass
system (c.m.s.) of initial particles the emitted photons are
concentrated manly in the narrow cones along momenta of each of
initial particles. The integral spectral cross section in the each
direction \cite{AB1, BFK1} with power accuracy (to within terms
$\sim m^2/\varepsilon^2$) is
\begin{equation}
d\sigma_c(1)=d\sigma_c(2)=\frac{4\alpha^3}{m^2}\frac{d\omega}{\omega}
\frac{\varepsilon'}{\varepsilon}\left(\frac{\varepsilon'}{\varepsilon}
+\frac{\varepsilon}{\varepsilon'}-\frac{2}{3}\right)
\left[\ln\left(\frac{4\varepsilon^2\varepsilon'}{m^2\omega}\right)
-\frac{1}{2}\right], \label{2}
\end{equation}
where $\varepsilon'=\varepsilon-\omega$. This cross section is the
largest which can be observed in colliding beam experiment and grows
logarithmically with energy, e.g. for $\varepsilon=1$~GeV and in the
interval $0.1 \leq \omega/\varepsilon \leq 1$ it attains $\sigma
\sim 10^{-25} {\rm cm}^2$. The main contribution to the cross
section gives the interval of low momentum transfer $q=\sqrt{-q^2}:
(m^3\omega/(4\varepsilon^2\varepsilon') \leq q \leq m)$ so that
deviation angle of radiating particle is less than $m/\varepsilon$.
When the scattering angle of an electron(positron) $\vartheta \gg
m/\varepsilon$ the radiation (within a logarithmic accuracy) is
directed along momenta of charged particles and photon emission
cross section from initial(i) and final(f) particle is (see
e.g.\cite{BFKh})
\begin{equation}
d\sigma_i(1)=\frac{\alpha}{\pi}\frac{d\omega}{\omega} \left(1
+\frac{\varepsilon^2}{\varepsilon'^2}\right)
\ln\left(\frac{\varepsilon\vartheta}{m}\right)d\sigma'_{e^+e^-},~
d\sigma_f(1)=\frac{\alpha}{\pi}\frac{d\omega}{\omega} \left(1
+\frac{\varepsilon'^2}{\varepsilon^2}\right)
\ln\left(\frac{\varepsilon\vartheta}{m}\right)d\sigma_{e^+e^-},
\label{3}
\end{equation}
where $d\sigma'_{e^+e^-}$ is the electron-positron scattering cross
section in the c.m.s. of final particles and
 $d\sigma_{e^+e^-}$ is the electron-positron scattering cross
section in the c.m.s. of initial particles.

\subsubsection{Double bremsstrahlung in electron-(electron)positron collision}

Radiation of two photons at $e^- - e^-(e^+)$ collision is of evident
interest for colliding beam experiments. The most interesting is the
case when photons are emitted in opposite directions along the
momenta of colliding particles, because the coincidence of two
photon registration permits to separate the effect from background.
This process was used as a monitor of beam collisions and for cross
sections normalization. The use of method of invariant integration
of tensors representing the radiation blocks mentioned in previous
subsection, simplifies essentially the calculation of integral
spectrum (the process is represented by 40 diagrams). The
qualitative properties of dependence of the process cross section on
momentum transfer given above for soft photons emission are valid
for any energy of photons, but the radiation blocks should be found
for hard photons. The spectrum of double bremsstrahlung in c.m.s.
has the form \cite{BG2}:
\begin{eqnarray}
\hspace{-10mm}&&d\sigma_{\omega_1\omega_2}=\frac{8\alpha^4}{\pi
m^2}\Bigg\{\left(1 +\frac{\omega_1}{\varepsilon}\right)\left(1
+\frac{\omega_2}{\varepsilon}\right)\eta_1 + \left[\left(1
+\frac{\omega_1}{\varepsilon}\right)\frac{\omega_2^2}{\varepsilon^2}+\left(1
+\frac{\omega_2}{\varepsilon}\right)\frac{\omega_1^2}{\varepsilon^2}\right]\eta_2
\nonumber \\
\hspace{-10mm}&&+\frac{\omega_1^2}{\varepsilon^2}\frac{\omega_2^2}{\varepsilon^2}\eta_3\Bigg\}
\frac{d\omega_1}{\omega_1}
\frac{d\omega_2}{\omega_2},~\eta_1=\nu(2)=\frac{5}{4}+\frac{7\zeta(3)}{8},~
\eta_2=\frac{1}{2}+\frac{7\zeta(3)}{8},~\eta_3=\frac{7\zeta(3)}{8}.
 \label{4}
\end{eqnarray}
Within a good numerical accuracy (better than 1\%) the expression in
curly brackets can be represented in the multiplicative
form:$\{\ldots\}=R(\omega_1)R(\omega_2)$, where
$R(\omega)=\sqrt{\eta_1}(1-\omega/\varepsilon)+\sqrt{\eta_3}\omega^2/\varepsilon^2$.
This form is very convenient for comparison with experimental data.
For soft photon the spectrum coincides with Eq.(\ref{12}).

The first observation of double bremsstrahlung was done in
Novosibirsk \cite{Gol}. A special study of the double bremsstrahlung
process as monitoring device was performed at ACO in Orsay
\cite{Aug6}. Achieved accuracy ($\sim 3\%$) was record for high
energy QED. It is striking that it was in measurement of 4-th order
process. Bearing in mind that the double bremsstrahlung can be
observed in quite clean conditions and has enough large cross
section which is known within very good accuracy, this process was
used as standard method for luminosity measurement in Novosibirsk,
Orsay and Frascati.

\subsubsection{Pair creation in electron-(electron)positron collision}

Another 4-th order process, which cross section doesn't decrease
with energy, is the {\it electroproduction} process $e^+e^-
\rightarrow e^+e^- + N$. There are two types of diagrams presenting
this process: 1) one-photon, where the final particles are created
by a photon radiated from one of lines of the initial electron or
positron; 2) two-photon, where the final particles are created at
collision of two photons, radiated from each of initial particles
(photon-photon colliding beams). The last mechanism is especially
important since the final states, including hadrons, which are even
at charge conjugation ($C=1$), can be produced with cross section
which doesn't decrease with energy, while in the one-photon channel
$C=-1$ and cross section of annihilation into hadron is decreasing
as $1/\varepsilon^2$.

The properties as well as values of contributions of one-photon and
two-photons diagrams differ significantly. The main contributions is
given by the two-photon diagrams. For creation of $e^+e^-$ pair in
electron-positron collision this contribution to the total cross
section is (with an accuracy up to terms $\sim m^2/\varepsilon^2$)
\begin{eqnarray}
&&\sigma_2=\frac{\alpha^4}{27\pi m^2}\Bigg[28 L^3 -178 L^2
+(490-82\pi^2)L+1203 \zeta(3) +\pi^2\left(78\ln
2+\frac{458}{3}\right)
\nonumber \\
&&-676\Bigg]=\frac{\alpha^4}{\pi
m^2}[1.04L^3-6.59L^2-11.8L+104]
 \label{20}
\end{eqnarray}
where $L=\ln 4\varepsilon^2/m^2$.  The main term ($\propto L^3$) was
found in 1934 by Landau and Lifshitz \cite{LL}, the rest of
logarithmic terms were calculated in \cite{BF3}, the constant was
calculated in \cite{KL}, see also review \cite{BFKK}.

Let us discuss this result.\\
1. In the limit $\varepsilon \gg m$ the cross section increases as a
cub of logarithm of energy.\\
2. Two of these logarithms originate from integration over the
transverse momenta of photons emitted from the initial particles,
the third one from
integration over the longitudinal momenta of the created pair.\\
3. At moderate energy the main term ($\propto L^3$) is compensated
essentially by the rest logarithmic terms and constant, e.g. for
$\varepsilon=5$~GeV the compensation diminishes the cross section
$\sigma_2$ Eq.(\ref{20}) which is about 2/3 of the main term.

The contribution to the total cross section of each set of
one-photon diagrams (connected with one line of initial particles)
is \cite{BF3}
\begin{equation}
\sigma_1=\frac{\alpha^4}{162\pi
m^2}(231\pi^2-2198)L=\frac{\alpha^4}{\pi m^2}0.51 L.
 \label{21}
\end{equation}
It is significantly smaller than $\sigma_2$.

If detectors measure outgoing particles at large polar angles only,
another kinematic region than in the main term of the cross section
$\sigma_2$ Eq.(\ref{20}) contributes in the corresponding cross
section. In the case when the both polar angle of created particles
are $\vartheta_+=\vartheta_-=\pi/2$ the differential over the angles
of created pair cross section of pair electroproduction has the form
\cite{BF2}
\begin{equation}
\frac{d\sigma}{dc_+dc_-d\varphi}=\frac{\alpha^4}{2\pi\varepsilon_0^2}
\frac{\ln\left(2\varepsilon^2(1-\cos\varphi)/m^2\right)}
{\sqrt{m^2/\varepsilon^2+2(1-\cos\varphi)}},
 \label{22}
\end{equation}
where $c_{\pm}=\cos \vartheta_{\pm},~\varphi=\varphi_+ -\varphi_-
+\pi$ is the non-coplanarity angle, $\varepsilon_0$ is the lowest
energy of particles of created pair (registration threshold). This
cross section has very sharp peak at $\varphi=0$. This important
peculiarity was used for observation. Large angle electroproduction
of electron-positron pair was first observed at VEPP-2 \cite{Bal4,
Bal5}. Data support the distribution Eq.(\ref{22}).

\subsection{Hadron production}

\subsubsection{Vector mesons}

One of the main goals of electron-positron colliders of the first
generation was production of pions and kaons to study
electromagnetic form factors of pions and kaons at the positive
(time-like) momentum transfers. In 1960, when the INP project was in
progress, Sakurai \cite{S} proposed the non-Abelian gauge theory of
strong interactions constructed upon the QED pattern. The gauge
invariance in QED means that the invariance under {\it local} phase
transformation $\psi\rightarrow \exp(ie\Lambda(x))\psi$ forces one
to introduce a new field, which is to be identified with the
electromagnetic field $A_{\mu}$ coupled universally (with the
constant $e$) to the conserved current constructed out of
electrically charged fields. To maintain the invariance under the
mentioned transformation it's necessary also to perform the
transformation $A_{\mu}\rightarrow A_{\mu}+\partial \Lambda/\partial
x^{\mu}$. In \cite{S} the Yang-Mills theory was used: if one
requires that the non-Abelian gauge transformation associated with
the isospin ${\bf I}$ conservation is local in character then one is
forced to introduce the vector field with the isospin ${\bf
I}=1~(\varrho^{\pm},\varrho^0$  mesons in modern notation) coupled
universally (with the constant $f_{\varrho}$) to the isospin current
constructed out of all fields having nonvanishing isospins.  In
\cite{S} this result was generalized by adding the baryon and
hypercharge conservation. This means appearance of two vector fields
($\omega, \phi$ neutral mesons in modern notation) coupled
universally (with the constants $f_{B},~f_{Y} $) to the baryon $B$
and hypercharge $Y$ currents constructed out of all fields having
nonvanishing baryon number (hypercharge). This development indicated
that one can hope for a first class physics at electron-positron
colliders.

Side by side with outlined above theory, the indications that
strong-interacting vector mesons play important role followed from
analysis of nucleon electromagnetic form factors and some inelastic
$\pi-p$ reactions. Connection between these approaches was
established by Gell-Mann and Zachariasen \cite{GZ}, where it was
stressed that in isovector electromagnetic form factors of hadrons
the diagrams dominate, where photon interacts with hadrons via
$\varrho^0$ meson. The model which takes into account only such
diagrams

\begin{figure}[h]
\begin{center}
\includegraphics[width=40mm]{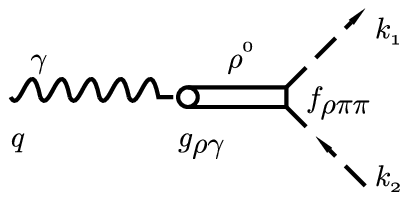}
\end{center}
\end{figure}
was called the {\it vector dominance model (VDM)}. In this model the
pion electromagnetic form factor is
$eF_{\pi}(t)=g_{\gamma\varrho}f_{\varrho\pi\pi}/(m^2-t)$, where
$g_{\gamma\varrho}$ is the amplitude of $\gamma \rightarrow
\varrho^0$ transition, $t=q^2$, $q$ is the photon momentum, $m$ is
the mass of $\varrho^0$ meson. Similarly, the isovector
electromagnetic form factor of nucleon is
$eF_{1N}(t)/2=g_{\gamma\varrho}f_{\varrho NN}/2(m^2-t)$. At zero
momentum transfer $F_{\pi}(0)=F_{1N}(0)=1$ because the electric
charge is universal. From
$g_{\gamma\varrho}f_{\varrho\pi\pi}/m^2=g_{\gamma\varrho}f_{\varrho
NN}/m^2=e$ it follows for all particles with isospin 1:
\begin{equation}
1) f_{\varrho\pi\pi}=f_{\varrho NN}=\ldots=f_{\varrho};\quad 2)
g_{\gamma\varrho}=em_{\varrho}^2/f_{\varrho}.
\label{h5}
\end{equation}
This is the consequence of $\varrho^0$ meson dominance in isovector
form-factor. What was done above can be expressed in the form
current-field identity:
\begin{equation}
j_{\mu}^{I_{\alpha}}(x)=m_{\varrho}^2\varrho_{\mu}^{\alpha}(x)/f_{\varrho},
\label{5}
\end{equation}
where $\varrho_{\mu}^{\alpha}(x)$ is the vector field describing
$\varrho$ mesons ($\alpha=1,2,3$).

The general form of the hadron electromagnetic current in the vector
dominance model, which is expressed in terms of $\varrho, \omega$
and $\phi$ mesons fields, reads \cite{KLZ}:
\begin{equation}
j_{\mu}^{had}(x)=e(m_{\varrho}^2\varrho_{\mu}^3(x)/f_{\varrho}+
m_{\omega}^2\omega_{\mu}(x)/f_{\omega}+m_{\phi}^2\phi_{\mu}(x)/f_{\phi}).
\label{5a}
\end{equation}
The field $\varrho_{\mu}^3$ is connected with isovector states (e.g.
$\pi^+\pi^-$), while the fields $\omega$ and $\phi$ are connected
with isoscalar states (e.g. $\pi^+\pi^-\pi^0, K^+K^-$) and can be
mixed up. The corresponding currents are \cite{S1}
\begin{eqnarray}
&& j_{\mu}^Y=\left[m^2_{\phi}\cos \vartheta_Y
\phi_{\mu}(x)-m^2_{\omega}\sin \vartheta_Y
\omega_{\mu}(x)\right]/f_Y,
\nonumber \\
&& j_{\mu}^B=\left[m^2_{\phi}\sin \vartheta_B
\phi_{\mu}(x)+m^2_{\omega}\cos \vartheta_B
\omega_{\mu}(x)\right]/f_B,
\label{6}
\end{eqnarray}
where $\vartheta_Y$ and $\vartheta_B$ are the mixing angles. The
hadron electromagnetic current is
$j_{\mu}^{had}=j_{\mu}^{3}+j_{\mu}^{Y}/2$. In the limit of exact
$SU(3)$ symmetry $\vartheta_B=\vartheta_Y=0,~
f_Y=\sqrt{3}f_{\varrho}/2$. In the broken $SU(3)$ symmetry
$\vartheta_B \neq 0,~\vartheta_Y \neq 0$, in mass mixing model
$\vartheta_B=\vartheta_Y=39^{\circ}$, in current mixing model
$\vartheta_B = 21^{\circ}, \vartheta_Y = 33^{\circ}$.

Taking into account that $\varrho^0$ meson is highly unstable, so
that $m=m_{\varrho}-i\Gamma_{\varrho}/2$.one has for $\varrho^0$
contribution to pion electromagnetic form factor
$F_{\pi}(t)=m_{\varrho}^2/(m_{\varrho}^2-t-i\Gamma_{\varrho}
m_{\varrho})$. This means that
\begin{equation}
|F_{\pi}(t)|^2 =
\frac{m_{\varrho}^4}{(m_{\varrho}^2-t)^2+\Gamma_{\varrho}^2
m_{\varrho}^2}\label{6a}
\end{equation}
 has sharp resonance peak at
$m_{\varrho}^2=t=4\varepsilon^2$ with the enhancement $\sim
m_{\varrho}^2/\Gamma_{\varrho}^2$. Similarly, the sharp peaks at
$t=m_{\omega}^2$ and $t=m_{\phi}^2$   should be observed in the
cross sections of production of isoscalar states (e.g.
$\pi^+\pi^-\pi^0, K^+K^-$).

Let us consider decay of vector meson into the electron-positron
pair and decay of $\varrho$ meson into the pion pair.
\begin{figure}[h]
\begin{minipage}[t]{80mm}
\begin{center}
\includegraphics[width=40mm]{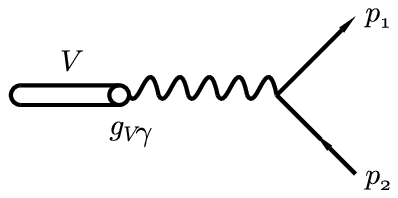}
\end{center}
\end{minipage}
\hfill
\begin{minipage}[t]{80mm}
\begin{center}
\includegraphics[width=40mm]{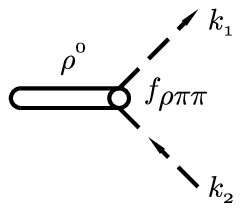}
\end{center}
\end{minipage}
\end{figure}

The partial width of decay $V \rightarrow e^+e^-$ is
\begin{equation}
\Gamma_{Ve^+e^-}=\frac{4\pi}{3}\frac{\alpha^2
m_V}{f_V^2}\left(1+\frac{2m_e^2}{m_V^2}\right)
 \left(1-\frac{4m_e^2}{m_V^2}\right)^{1/2}=\frac{4\pi}{3}\frac{\alpha^2
m_V}{f_V^2}\left[ 1+O\left(\frac{m_e}{m_V}\right)^4\right],
\label{7}
\end{equation}
where $m_e$ is the electron mass, the equality
$g_{V\gamma}=em_V^2/f_V$ is used. Since the value $\Gamma_{Ve^+e^-}$
is measured quite accurately, this expression can be used for
determination of constants $f_V$. The constant $f_{\varrho\pi\pi}$
is determined from the width of decay $\rho\rightarrow\pi\pi$
\begin{equation}
\Gamma_{\rho\pi\pi}=\frac{f_{\varrho\pi\pi}^2}{4\pi}\frac{m_{\varrho}
}{12}v_R^{3}, \label{8}
\end{equation}
where $v_R=\sqrt{1-4\mu^2/m_{\rho}^2}$ is the velocity of the
created pion, $\mu$ is the pion mass.

The total cross section of $e^+e^-$ annihilation into pair of
pseudoscalar ($\pi, K$) particles is
\begin{equation}
\sigma=\frac{\pi\alpha^2}{3t}v^{3}|F(t)|^2,\label{9}
\end{equation}
where $v=\sqrt{1-4\mu^2/t}$ and $\mu$ are the velocity and mass of
produced particle, $F(t)$ is the electromagnetic form factor of
corresponding particle, $t=4\varepsilon^2$. So at the resonance
energy $m_{\varrho}^2=t$ one has the cross section
$\sigma_R=(12\pi/m_{\varrho}^2)(\Gamma_{\varrho\pi\pi}
\Gamma_{\varrho e^+e^-}/\Gamma_{\rho}^2)$. This formula has a
transparent meaning, since in the quantum theory the resonance cross
section in the channel with angular momentum $J$ is
$\sigma_R=\pi\lambda^2(2J+1)\Gamma_i\Gamma_f/\Gamma_V^2=
4\pi(2J+1)/m_V^2(\Gamma_i\Gamma_f/\Gamma_V^2)$, where
$\lambda=1/\varepsilon=2/m_V,~\Gamma_{i(f)}$ is the width of the
resonance into channel $i(f)$, $\Gamma_V$ is the total width.

The transition of photon into the vector meson (which is the
contribution to the hadronic polarization of vacuum) can appear in
the purely QED processes such as $e^-e^+$ elastic scattering or the
conversion process $e^-e^+\rightarrow\mu^-\mu^+$. The last reaction
is more appropriate since only one (annihilation) diagram
contributes. The most pronounced effect will be near resonance
$2\varepsilon \simeq m_V$. The cross section of the
$e^-e^+\rightarrow\mu^-\mu^+$ process (see Eq.(\ref{1})) with the
transition $\gamma V$ taken into account acquires an additional
factor \cite{BG4}
\begin{equation}
|1+\frac{g^2}{\alpha}~\frac{m_V}{2(2\varepsilon-m_V)+i\Gamma_V}|^2,
 \label{8aa}
\end{equation}
where $g$ is the effective coupling constant $Ve^-e^+$ or
$V\mu^-\mu^+$. It can be expressed in terms of branching ratio
$B_{Ve^-e^+}=\Gamma_{Ve^-e^+}/\Gamma_V$ and $B_{V\mu^-\mu^+}$, see
Eq.(\ref{7}). The prediction \cite{BG4} was made for only known in
1963 narrow $\omega$ meson. The factor Eq.(\ref{8aa}) results in
oscillation of the process cross section with respect to the QED
prediction in the narrow energy interval (the width $\sim \Gamma_V$)
near $\varepsilon \simeq m_V/2$: first the cross section is going
down, than it turns up and crosses the prediction very close to
$\varepsilon = m_V/2$, attains some maximal value and than returns
to the prediction.

Since the ratio  $\Gamma_{\varrho}/m_{\varrho}$ turns out to be not
very small the corrections $\propto \Gamma_{\varrho}/m_{\varrho}$
and dependence of term with $\Gamma_{\varrho}$ in the resonance
denominator $m_{\varrho}^2-t-i\Gamma_{\varrho} m_{\varrho}$ on the
pion momentum becomes significant \cite{GS}: $\Gamma_{\varrho}
m_{\varrho} \rightarrow \Gamma_{\varrho}(p/p_0)^3
m_{\varrho}^2/2\varepsilon $, where $p$ is the momentum of created
pion and $p_0$ is the momentum at $m_{\varrho}=2\varepsilon$. With
regard for final width of $\varrho$ meson Eq.(\ref{h5}) is modified:
$f_{\varrho\pi\pi}^2=1.15~ f_{\varrho}^2$. Besides, since the masses
of $\varrho^0$ and $\omega^0$ mesons appears to be very close, the
contribution of process $e^+e^- \rightarrow \omega^0 \rightarrow
\pi^+\pi^-$ (the $\varrho-\omega$ interference) should be taken into
account, in spite of the fact that in the mentioned channel the
isospin invariance is violated, because at resonance the cross
section $\propto (m_V/\Gamma_V)^2$ and $\Gamma_{\omega} \ll
\Gamma_{\varrho}$. As a result the $\varrho$-meson excitation curve
becomes asymmetric.

The first indications on existence of the vector mesons was obtained
in the hadronic reactions in 1961. The particle which is now called
$\varrho$ meson was observed in inelastic $\pi p$ collisions
\cite{St} with mass $m_{\varrho}$ in interval 700-770 MeV and width
$\Gamma_{\varrho}\sim 90$~MeV. The $\omega$ was seen in reaction
$\bar{p}p \rightarrow 2\pi^+\pi^-\pi^0$ \cite{MA} with mass
$m_{\omega}$=787 MeV and $\Gamma_{\omega} < 30$~MeV. Extraction of
properties of the vector mesons in the hadronic reaction is quite
ambiguous due to involvement of strong interactions and only
electron-positron colliders permitted to perform full scale study of
the vector mesons.

\begin{figure}[h]
\begin{center}
\includegraphics[width=50mm]{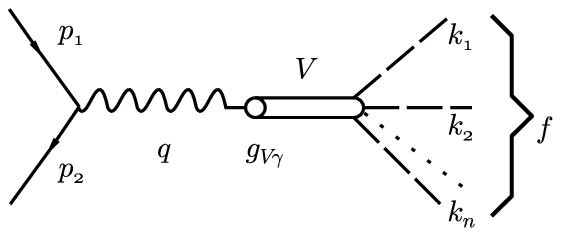}
\end{center}
\end{figure}

The neutral vector mesons at colliding $e^-e^+$ beams were observed
first at VEPP-2 storage ring in INP, Novosibirsk in 1967 \cite{Au},
\cite{Au1}, where the cross section of production of the
$\pi^-\pi^+$ pair was measured in the region of $\varrho$-resonance
and the excitation curve was obtained. Later the same measurement
was performed at ACO storage ring in LAL, Orsay \cite{Aug2},
\cite{Aug1}, \cite{Ben}.

In isoscalar channel production of $\pi^-\pi^+\pi^0$ was observed
first at ACO storage ring in LAL, Orsay  in the region of
$\omega$-resonance \cite{Aug3} and the region of $\phi$-resonance
\cite{Aug4}. The last channel was observed also in INP, Novosibirsk
\cite{Bal3}. The reactions $\phi \rightarrow K_L^0 K_S^0, K^+ K^-$
were observed at ACO storage ring in LAL, Orsay \cite{Aug4},
\cite{Bi} and at VEPP-2 storage ring in INP, Novosibirsk
\cite{Bal3}. In these experiments the excitation curves were
measured and resonance parameters were obtained.

Radiative modes of decay of $\omega$ and $\phi$ mesons into
$\eta\gamma,~ \pi^0\gamma,~\pi^+\pi^-\gamma$ were investigated in
LAL, Orsay \cite{Ben1}.  The multi-hadron production in
electron-positron annihilation was discovered at VEPP-2 storage ring
in INP, Novosibirsk \cite{KOS}. The production of
$\pi^+\pi^-,~K^+K^-$ pairs in electron-positron annihilation at
energy higher than $\phi$ resonance mass was observed  at VEPP-2
storage ring in INP, Novosibirsk \cite{Bal6}. The vacuum
polarization in the process $e^-e^+\rightarrow\mu^-\mu^+$ due $\phi$
meson contribution according to Eq.(\ref{8aa}) (where
$g^2=3B,~B=\sqrt{B_{\phi e^-e^+}B_{\phi \mu^-\mu^+}}$) was observed
in LAL, Orsay \cite{Aug5}, the magnitude of oscillation was $\sim
10\%$. The calibration of storage ring energy was performed using
the angular distribution of pions in reaction $e^+e^-
\rightarrow\phi \rightarrow K_L^0 K_S^0, K_S^0\rightarrow
\pi^+\pi^-$ \cite{Bal7}.

For review see e.g. \cite{Ba1}, \cite{PR}.

The results obtained at the electron-positron colliders confirmed
the basic predictions of the vector dominance model, which appears
to be remarkably successful, and become outstanding achievement of
the new method.

The recent parameters of vector mesons are given in the Table 1
below. These parameters differs from measured in cited above
experiments 1967-1972 on the level of one standard deviation but
here the accuracy is improved significantly.

\begin{table}[h]
\begin{center}
{\sc Table 1}~ {Parameters of vector mesons (PDG 2004)}
\end{center}
\begin{center}
\begin{tabular}{*{6}{|c}|}
\hline meson &$m_V$(MeV)&$\Gamma_V$(MeV)&$\Gamma_{Ve^+e^-}$(keV)&$f_V^2/4\pi$&$g_{V\gamma}({\rm GeV}^2)$ \\
\hline $\varrho$&775.8$\pm$ 0.5&146.4$\pm $1.5&7.02$\pm $0.11&1.96$\pm$ 0.03&0.121$\pm $0.001\\
\hline $\omega$&782.59$\pm$ 0.11&8.49$\pm $0.08&0.60$\pm $0.02&23.2$\pm$ 0.8&0.036$\pm $0.001\\
\hline $\phi$&1019.456$\pm$ 0.020&4.26$\pm $0.05&1.27$\pm $0.04&14.2$\pm$ 0.4&0.078$\pm $0.001\\
\hline
\end{tabular}
\end{center}
\end{table}

For these parameters one has using Eq.(\ref{8})
$f_{\varrho\pi\pi}^2/4\pi=2.79 \pm 0.03$.

Just by that time when the main results obtained at the first
generation of electron-positron colliders were published
(1972-1973), the quantum chromodynamics (QCD), which is the
non-Abelian gauge theory, emerged \cite{GW, Po} and in a short time
was accepted as a strong interaction theory. In QCD the basic
components are quarks and gluons. In this theory the vector mesons
discussed above are the composite systems each consisting of light
($u, d, s$) quark and antiquark with parallel spins coupled by the
gluon field (e.g. the state of $\varrho$ meson is
$\varrho=(u\bar{u}+d\bar{d})/\sqrt{2}$). For this picture the VDM is
an effective theory valid for energies up to $ \sim 1~$ GeV. Since
the parameters of vector mesons are now measured within percent
accuracy the deviations from exact VDM are seen (e.g. from
parameters given above the ratio
$f_{\varrho\pi\pi}^2/f_{\rho}^2$=1.42 and not 1.15). Description of
vector mesons in QCD frame for mentioned parameters lies indeed in
region of strong coupling and should be done in scope of
non-perturbative methods. Such analysis should not only explain the
origin of the VDM but also clarify deviations from exact VDM. In
lattice QCD the recent progress is on the level of $\varrho$ meson
mass calculation \cite{Ar}. Since there is no other reliable
methods, the vector dominance is still a challenge for QCD.

\subsubsection{Radiative return}

The cross section of the process passing through the vector meson
contains the resonant factor of the type of Eq.(\ref{6a}). Because
of this the radiative corrections to the cross section of such
process, considered first by Victor Fadin and I \cite{BF}, depended
strongly on energy. This is a consequence of photon emission from
initial particles, which leads to decrease of produced particle
energy.

\begin{figure}[h]
\begin{center}
\includegraphics[width=50mm]{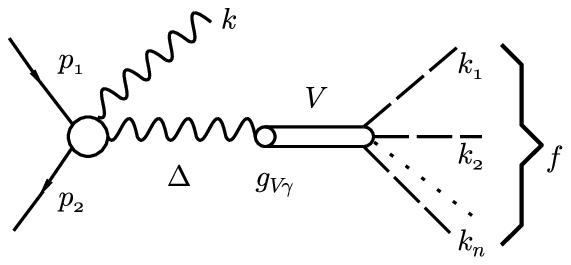}
\end{center}
\end{figure}

If the initial particles energy is higher than the resonance one
$\varepsilon>\varepsilon_R=m_V/2$, "then the initial particle
radiation can "turn" the cross section back to the resonance (when
final particles energy in their c.m.s. is equal to the resonant
one). Since the resonance cross section essentially exceeds the
cross section far away from resonance, this leads to a fast increase
of the radiative corrections at $\varepsilon>\varepsilon_R$".

Within the logarithmic accuracy in the case of soft photons the
radiative corrections, given only by the initial electron and
positron, calculated in \cite{BF} are written as
\begin{eqnarray}
&& d\sigma=d\sigma_0(1+\delta(\varepsilon)), \quad
\delta(\varepsilon)=\frac{\alpha}{\pi}\Bigg\{ (2\ln
2\gamma-1)\Bigg[\ln
\frac{\omega^2[\tau(0)^2+\Gamma_V^2]}{\varepsilon^2
[\tau(\omega)^2+\Gamma_V^2]}
\nonumber \\
&& + \frac{2\tau(0)}{\Gamma_V}\arctan\left(\frac{2\omega\Gamma_V}
{\tau(\omega)\tau(0)+\Gamma_V^2} \right) \Bigg]+\frac{13}{3}\ln
2\gamma \Bigg\}, \label{9a}
\end{eqnarray}
where $d\sigma_0$ is the process cross section without radiative
corrections depending on final particles momenta, $\varepsilon$ is
the initial energy of electron(positron) in their c.m.s.,
$\gamma=\varepsilon/m_e$, $\omega$ is the maximal permissible by the
event selection energy of photon emitted in the direction of initial
particle, $\tau(\omega)=2(2\varepsilon-m_V-\omega)$, note that the
square of invariant mass of final system is
$\Delta^2=4\varepsilon(\varepsilon-\omega)$. This formula can be
applied not far from resonance \footnote{The process $e^+e^-
\rightarrow \pi^+\pi^-\gamma$ was analyzed in \cite{BKho}, where the
exact in Born approximation and very compact expression for the
integral spectrum in terms of $\Delta^2$ was calculated.}. The above
effect is determined mainly by the second term in the square
brackets Eq.(\ref{9a}). It may turn out that the term
$\tau(0)/\Gamma_V \gg 1$ and its factor is of the order of 1. Then
it is possible that $4\alpha \tau(0)\ln 2\gamma/\pi \Gamma_V \geq 1$
or $\delta(\varepsilon) \geq 1$! Such paradoxical situation arose
due to the fact that cross section of the process with photon
emission by the initial particle turns out to be larger than the
cross section of elastic process (without inclusion of radiative
corrections) at a given energy of the initial particles. The
increase of $\delta(\varepsilon)$ stops when the condition of event
selection forbids emission of photon with an energy sufficient for
the shift to the resonance.

It should be noted that Eq.(\ref{9a}) can be applied for an
arbitrary process of particle production passing through the
resonant state.

The radiative return method basing on equations of the type
Eq.(\ref{9a}) is widely used now at meson factories (BELLE, BABAR,
CLEO-C, DAPHNE) for study of particular hadronic reactions from
production energy threshold up to the energy close to the machine
energy $2\varepsilon$ (for recent review see e.g. \cite{C}). The
behavior of reaction $e^+e^- \rightarrow p\bar{p}$ ($p$ is a proton)
near threshold, process $e^+e^- \rightarrow 3\pi$ for energy
$\varepsilon
> 0.7$~GeV (higher than operational energy of VEPP-2) are among the results
obtained.

\subsection{Polarization}

\subsubsection{Radiative polarization of electrons in storage rings}

During extended motion in a magnetic field electrons and positrons
can be polarized as a result of photon emission. The polarization
arises because the probability of radiative transition with spin
flip depends on the orientation of the initial spin. Existence of
this mechanism was pointed out by Ternov, Sokolov et al. \cite{TLK},
\cite{ST}. The solution of Dirac equation in the uniform magnetic
field was used in these paper. However, it is known that typical
conditions accelerators correspond to very high quantum numbers,
e.g. for $H \sim 10^4~Oe$ and energy $\sim 1$ GeV the main quantum
number $\sim 10^{15}$. This means that the motion of particle in
accelerator and storage ring is almost classical. We with Valery
Katkov developed an operator method for investigation of spin
phenomena\footnote{Later the general quasiclassical operator method
was developed by Katkov and I, which is actually the formulation of
QED in an arbitrary electromagnetic field at high energy \cite{BK2},
\cite{BK3}.} in a quasiclassical approximation \footnote{Similar
procedure was used by Schwinger\cite{Sw} to find the first quantum
correction to the intensity of electron radiation in a magnetic
field.}. The total probability of spin-flip radiative transition per
unit time valid in an arbitrary magnetic field is \cite{BK1}
\begin{equation}
W^{\zeta}=\frac{1}{2T}\left[1-\frac{2}{9}(\mbox{\boldmath$\zeta$}\textbf{v})^2
-\frac{8\sqrt{3}}{15|\dot{\bf v}|}(\mbox{\boldmath$\zeta$}(\dot{\bf
v}\times \textbf{v}))\right],\quad
\frac{1}{T}=\frac{5\sqrt{3}}{8}\frac{\alpha \gamma^5 |\dot{\bf
v}|^3}{m^2},
\label{10}
\end{equation}
where $\mbox{\boldmath$\zeta$}=\mbox{\boldmath$\zeta$}(t)$ is the
unit spin vector of an electron, $\textbf{v}$ and $\dot{\bf v}$ are
the velocity and acceleration of an electron, $T$ is the
characteristic time of polarization\footnote{For magnetic radius
$r=150$~cm (VEPP-2 facility) and $\varepsilon=700$~MeV one has
$T=38$~minutes.}. For the longitudinal polarization
$(\mbox{\boldmath$\zeta$}(\dot{\bf v}\times \textbf{v}))= 0$ the
remaining terms $1-(2/9)(\mbox{\boldmath$\zeta$}\textbf{v})^2$ do
not depend on whether the spin is directed along or opposite to the
velocity, so that the radiation does not change the spin states with
longitudinal polarization. A different situation arises in the case
of transverse polarization $(\mbox{\boldmath$\zeta$}\textbf{v})=0$.
In this case the transition probability depends on the spin
orientation. For electrons ($e < 0$) the probability of a transition
from a state with spin along the field to a state with spin opposite
to the field is higher than the probability of the inverse
transition. For positrons ($e > 0$) the opposite situation occurs.
Thus, the resulting polarization (radiative polarization) is
transverse and for electrons is directed opposite to the field and
for positrons along it.

It is very important that the probability Eq.(\ref{10}) is given in
the same terms as used in the quasiclassical equation for spin
motion of Bargmann-Michel-Telegdi (BMT)  in an external field
\cite{BMT}. The point is that the radiative polarization is rather
slow process which evolves at background of rather complicate spin
motion (described by BMT equation) in a storage ring. The kinetic
equation which takes into account both factors was derived by Valery
Katkov, Volodya Strakhovenko and I \cite{BKS1}, \cite{BKS2}:
\begin{eqnarray}
&& \frac{d\mbox{\boldmath$\zeta$}}{dt}=\frac{e}{\varepsilon}
\left(\mbox{\boldmath$\zeta$}\times \left(\mu
\textbf{H}_R+\textbf{H}_E\right)\right)-\frac{1}{T}
\left[\mbox{\boldmath$\zeta$}-\frac{2}{9}\textbf{v}(\mbox{\boldmath$\zeta$}\textbf{v})
-\frac{8\sqrt{3}}{15|\dot{\bf v}|}(\dot{\bf v}\times
\textbf{v})\right],
\nonumber \\
&&
\textbf{H}_R=\gamma\left[\textbf{H}-\frac{\textbf{v(vH)}}{1+1/\gamma}
-(\textbf{v}\times\textbf{E})\right],~
\textbf{H}_E=\textbf{H}-\frac{(\textbf{v}\times
\textbf{E})}{1+1/\gamma} ,\label{13}
\end{eqnarray}
where $\mu=\alpha/2\pi$ is the anomalous magnetic moment of an
electron, $\textbf{E}$ and $\textbf{H}$ are the fields in the
laboratory system, $\textbf{H}_R$ is the magnetic field in the rest
system of the electron. The first term in this equation is just BMT
equation, while the second term appearing due to the spin-flip
transitions leads to variation of $|\mbox{\boldmath$\zeta$}|$.

In the simplest case of circular motion in a homogeneous magnetic
field decomposing the vector $\mbox{\boldmath$\zeta$}$ over the unit
vectors ${\bf e}_1={\bf v}/|{\bf v}|,~ {\bf e}_2=\dot{\bf
v}/|\dot{\bf v}|$ and ${\bf e}_3=({\bf e}_1\times {\bf e}_2)$ one
has from the above equation
\begin{equation}
\dot{\zeta_1}=-\frac{7}{9}\frac{\zeta_1}{T}-\Omega \zeta_2,\quad
\dot{\zeta_2}=\Omega \zeta_1-\frac{\zeta_2}{T},\quad
\dot{\zeta_3}=-\frac{1}{T}\left(\zeta_3+\frac{8}{5\sqrt{3}}\right),
\label{13a}
\end{equation}
where $\Omega=\mu\gamma|\dot{\bf v}|$. The solution of this set is
\begin{equation}
\zeta_{\perp}(t)=\zeta_{\perp}(0)\exp\left(-\frac{8t}{9T}\right),\quad
\zeta_3(t)=-\frac{8}{5\sqrt{3}}+\left(\zeta_3(0)+\frac{8}{5\sqrt{3}}\right)
\exp\left(-\frac{t}{T}\right), \label{13b}
\end{equation}
where $\zeta_{\perp}(t)=\sqrt{\zeta_1^2(t)+\zeta_2^2(t)}$, it was
taken into account that $\Omega \gg 1/T$. So the spin rotates around
the ${\bf e}_3$ axis, the transverse component decays during a time
$\sim T$, while the nondecaying term -$8/5\sqrt{3}$ in $\zeta_3$
gives a finite polarization ($\sim 0.924$) which does not depend on
the initial value of the vector $\mbox{\boldmath$\zeta$}$. The
polarization is oriented along the vector  $(\dot{\bf v} \times {\bf
v})$.

Side by side with outlined development the very important result
concerning behavior of spin vector was obtained by Derbenev,
Kondratenko and Skrinsky \cite{DKS}. It was shown that the stable
direction of polarization exists for solution of BMT equation
($\textbf{n}(t)=\textbf{n}(t+\tau)$, $\tau$ is the period of
revolution) for closed orbits in storage ring with arbitrary field.

Basing on mentioned above results and analysis of depolarization
effects\cite{DKS1} Derbenev and Kondratenko obtained the following
equation for the equilibrium degree of polarization for the time
essentially larger than $T$ \cite{DK}
\begin{equation}
\mbox{\boldmath$\zeta$}{\bf n} \equiv \zeta_{\bf
n}=-\frac{8}{5\sqrt{3}}\frac{<|\dot{\bf v}|^2({\bf v} \times\dot{\bf
v})({\bf n}-\gamma\frac{\partial{\bf n}}{\partial
\gamma})>}{<|\dot{\bf v}|^3[1-\frac{2}{9}({\bf n}{\bf v})^2 +
\frac{11}{18}(\frac{\partial{\bf n}}{\partial \gamma})^2]>},
\label{14}
\end{equation}
where $<\ldots>$ means averaging over azimuth and particle ensemble
in storage ring. This formula summarize many contributions: the
external factor $8/5\sqrt{3}$ was found by Sokolov and Ternov
\cite{ST}, the terms $<|\dot{\bf v}|^2({\bf v} \times\dot{\bf
v}){\bf n}>$ and $<|\dot{\bf v}|^3[1-\frac{2}{9}({\bf n}{\bf
v})^2]>$ follow directly from Eq.(\ref{13}), the term
$\gamma\frac{\partial{\bf n}}{\partial \gamma}$ reflecting
perturbation of quantization axis ${\bf n}$ due to influence of
spin-dependent part of the magnetic bremsstrahlung is the invention
of Derbenev and Kondratenko \cite{DK}, the term
$\frac{11}{18}<(\frac{\partial{\bf n}}{\partial \gamma})^2>$
describes electron beam depolarization due to chaotic jumps of the
trajectory because of quantum nature of radiation process discovered
by  me and Yuri\"{\i} Orlov during his short stay in Novosibirsk
\cite{BO}.  Emerging, conservation, manipulation and measurement of
radiative polarization are discussed in detail in \cite{Ba2}, see
also Sec.14 in \cite{BKF}.

\subsubsection{Measurement of electron polarization}

\hspace{3mm} I.~~{\it High energy processes}
\vspace{1mm}
\newline We have shown with Victor Fadin that the cross sections
of two-particle production at electron-positron annihilation are
extremely sensitive to electron and positron polarizations
\cite{BF1}. So, these reactions can be used for polarization
measurement.

The cross section for production of a pair of pseudoscalar particles
($\pi^+\pi^-,~K^+K^-$, ~$K_S^0K_L^0$) in annihilation of
transversely (and antiparallel) polarized electrons and positrons
has the form
\begin{equation}
\sigma_{2p}(\vartheta,
\varphi)=\sigma_{2p}^0(\vartheta)[1-|\mbox{\boldmath$\zeta$}_1|
|\mbox{\boldmath$\zeta$}_2|\cos 2\varphi],
 \label{15}
\end{equation}
where $|\mbox{\boldmath$\zeta$}_1|$ and
$|\mbox{\boldmath$\zeta$}_2|$ are the degrees of polarization of the
positrons and electrons, $\varphi$  is the angle between the plane
of production (the plane passing through the momenta of the initial
particle {\bf p} and the final particle {\bf q}) and the plane
perpendicular to the spin direction (the plane of the orbit),
$\sigma_{2p}^0(\vartheta)$ is the cross section for unpolarized
particles:
$\sigma_{2p}^0(\vartheta)=\alpha^2v^3\sin^2\vartheta|F(t)|^2/8t$ (cf
with Eq.(\ref{9})), $\vartheta$ is the angle between {\bf p} and
{\bf q}. . If the initial particles are completely polarized
 $|\mbox{\boldmath$\zeta$}_1|=|\mbox{\boldmath$\zeta$}_2|=1$,
then $\sigma_{2p}(\vartheta, \varphi=0)=0$ (the production plane
coincides with the orbit plane) and $\sigma_{2p}(\vartheta,
\varphi=\pi/2)=2\sigma_{2p}^0((\vartheta)$ (the production plane is
perpendicular to the orbit plane, so that the spin vector lies in
the production plane).

For production of a pair of muons one has
\begin{equation}
\sigma_{2\mu}(\vartheta,\varphi)=\frac{\alpha^2}{4t}v[2-v^2\sin^2
\vartheta[1-|\mbox{\boldmath$\zeta$}_1|
|\mbox{\boldmath$\zeta$}_2|\cos 2\varphi]].
 \label{16}
\end{equation}
For relativistic muons $v \simeq 1$, and we have for completely
polarized particles $\sigma_{2\mu}(\vartheta=\pi/2,\varphi=\pi/2)=0$
(muon momentum directed along the spin) and
$\sigma_{2\mu}(\vartheta=\pi/2,\varphi=0)=2\sigma_{2\mu}^0(\vartheta)$
(muon momentum perpendicular to the spin). \vspace{3mm}

\hspace{3mm}II.~~{\it Internal scattering effects and polarization
measurement}~\cite{BKho1}
\vspace{1mm}
 \newline It is well known that an important cause of the loss
of particles in storage ring is the electron-electron scattering
inside the bunch \cite{BCG}. If this scattering occurs into a
sufficiently large angle and is such that particles with a large
transverse momentum and small longitudinal momentum (in the rest
system of the beam) acquire a large longitudinal moment, then in
conversion to the laboratory system the longitudinal momentum is
subject to the relativistic transformation and can turn out to be
larger than the permissible value. As a result the particles are
lost. Under some conditions the lifetime of a beam in a storage ring
is determined just by the Touschek effect. Internal scattering
effects depend on the particle polarization, since the
electron-electron scattering cross section at the large angles which
determine the internal scattering effect depends substantially on
electron polarization. The beam lifetime $\tau$  ($\tau$ is the time
in which the number of particles decreases by a factor of two)is
determined by the coefficient $\alpha_b$:$1/\tau=\alpha_b N_0,~ N_0$
is the initial number of particles in the beam. For example, for a
Gaussian distribution of radial momenta of the electrons in the beam
one has
\begin{eqnarray}
&&\alpha_b=\frac{2\sqrt{\pi}\alpha^2 m}{V(\Delta p)^2\delta
q}\Bigg[\ln\frac{2\varepsilon}{\Delta
p}-\frac{7}{4}-\frac{\mbox{\boldmath$\zeta$}_1
\mbox{\boldmath$\zeta$}_2}{4} + 2\sqrt{\pi}\frac{\delta
q}{m}\exp\left(\frac{m^2}{\delta
q^2}\right)\left(1+\frac{m^2}{2\delta q^2}\right)
\nonumber \\
&&\times \left(1-\Phi\left(\frac{m}{\delta q}\right)\right)
-\sqrt{\pi}\int_{0}^{\frac{m}{\delta q}}e^{x^2}(1-\Phi(x))dx\Bigg]
,\label{16a}
\end{eqnarray}
where $V$ is the volume of the beam in the laboratory system,
$\Delta p$ is the maximum permissible deviation of momentum from the
equilibrium value in the laboratory system, $\delta q$ is the mean-
square value of the momentum distribution, $\varepsilon$ is the
electron energy in the laboratory system,
$\mbox{\boldmath$\zeta$}_{1,2}$ are the polarization vectors of
electrons in the bunch, $\Phi(x)$ is the probability integral.
 This dependence of the internal scattering effect on polarization
is used to measure the polarization of electrons in a storage ring.
\vspace{3mm}

\hspace{3mm} III.~~{\it Measurement of polarization by means of
Compton scattering}~\cite{BKho2}
\vspace{1mm}
 \newline In Compton scattering of circularly polarized photons
by transversely polarized high-energy electrons, terms in the cross
section arise which depend on the electron polarization vector. In
head-on collisions of laser photons (with energy $\omega_1$) with
high-energy electrons, the final photons are emitted mainly in a
narrow cone with an angle $\sim 1/\gamma$ relative to the initial
electron direction. The cross section can be written in a form
\begin{equation}
d\sigma=d\sigma_0+d\sigma_1|\mbox{\boldmath$\zeta$}_1|
\xi_2\sin\varphi,
 \label{17}
\end{equation}
where $d\sigma_0$ is the cross section for unpolarized particles,
$\xi_2$ is the degree of circular polarization of the photons, and
$\varphi$ is the angle between the plane perpendicular to the vector
$\mbox{\boldmath$\zeta$}_1$ and the scattering plane. The azimuthal
asymmetry coefficient has the form
\begin{equation}
P=\frac{d\sigma_1}{d\sigma_0}=-\frac{2\lambda
n(1+n)^2}{2\lambda^2(1+n^2)+(1+n^2+2\lambda)(1+n^4)},
 \label{18}
\end{equation}
where $\lambda=2\omega_1\varepsilon/m^2$, the photon scattering
angle measured from the direction of electron momentum is written as
$\vartheta=n/\gamma \ll 1$.

\newpage

\hspace{3mm} IV.~~{\it The first experiment}
\vspace{1mm}

The first experimental study of the radiative polarization of
electrons has been carried out in the storage ring VEPP-2 in INP,
Novosibirsk (see \cite{Ba2}). The polarization measurement was
accomplished by the method described above in the paragraph II,
which utilizes the dependence of internal scattering effects on the
polarization of the electrons in the bunch (see Eq.(\ref{16a})). For
the energy chosen ($\varepsilon$ = 650 MeV) the polarization time is
$T \simeq$ 50 min and the theoretical degree of polarization during
the experiment is $|\zeta_3(2T)|\simeq 0.80$ (see Eq.(\ref{13b})).
In this experiment it was extremely important to exclude the effect
of depolarizing factors. For this purpose it is necessary first of
all to be sufficiently far from spin resonances. If the depolarizing
effects are taken into account, then the expected degree of
radiative polarization is $|\zeta_3^{th}(2T)|\simeq 0.66$.

The measurements were made in the following way. The electron beam
in the storage ring was polarized for a time $t \simeq 2T$, and the
particles leaving the beam as a consequence of internal scattering
effects were recorded by two counters. Then the beam was depolarized
by application of an external longitudinal field. In this case the
rate of departure of particles from the beam increases (i.e., the
number of counts in the counters increases). The experimental
results was obtained for an energy $\varepsilon=638.8\pm0.8~$ MeV. A
jump was seen in the counting rate, occurring at the turning on of
the depolarizing field. From the size of the jump one can deduce the
following value of the degree of polarization of the electron beam:
\begin{equation}
|\zeta_3^{exp}(2T)|\simeq 0.52\pm 0.13,
 \label{19}
\end{equation}
 which is consistent with the expected
value of the degree of polarization given above with inclusion of
depolarizing effects $|\zeta_3^{th}(2T)|\simeq 0.66$, although it is
somewhat smaller. This was the first experimental proof of the
existence of radiative polarization.

\section{Conclusion}

Let us list the main results obtained at the electron-positron
colliders of the first generation.

At the electron-positron colliding beam facility VEPP-2 in INP,
Novosibirsk (the maximal observed luminosity
$L=3\times10^{28}~cm^{-2}s^{-1}$):

\hspace{3mm} 1. The first observation of hadron production at
electron-positron collider (1967), study of $\varrho$ meson.

\hspace{3mm} 2. The first observation of two-photon annihilation
($e^+e^-\rightarrow 2\gamma$).

\hspace{3mm} 3. The first observation and study of the radiative
polarization of beam in storage ring.

\hspace{3mm} 4. The first observation and study of the two-photon
process (production of additional electron-positron pair).

\hspace{3mm} 5. Check of~ QED at $e^+e^-$ collision.

\hspace{3mm} 6. Check of~ QED in reaction $e^+e^-\rightarrow
\mu^+\mu^-$.

\hspace{3mm} 7. Systematic study of $\varrho,\omega,\phi$ mesons.

\hspace{3mm} 8. Discovery of the multi-hadron production in
electron-positron annihilation.

\hspace{3mm} 9. Study of production $\pi^+\pi^-,~K^+K^-$ pairs in
electron-positron annihilation at energy higher than $\phi$
resonance mass.\\

At the electron-positron colliding beam facility ACO in LAL, Orsay
(the maximal observed luminosity $L=10^{29}~cm^{-2}s^{-1}$):\\

\hspace{3mm} 1. The first observation and study of $\omega$ meson.

\hspace{3mm} 2. The first observation and study of $\phi$ meson.

\hspace{3mm} 3. Study of $\varrho$ meson.

\hspace{3mm} 4. Study of $\phi-\omega$ and $\varrho-\omega$
interference.

\hspace{3mm} 5. Study of radiative modes of decay of $\omega$ and
$\phi$ mesons into $\eta\gamma,~ \pi^0\gamma,~\pi^+\pi^-\gamma$.

\hspace{3mm} 6. Study of $\mu$ meson pair creation
($e^+e^-\rightarrow \mu^+\mu^-$).

\hspace{3mm} 7. Check of QED at $e^+e^-$ collision.

\hspace{3mm} 8. Study of vector dominance model.

\hspace{3mm} 9. Observation of $\phi$ meson contribution to vacuum
polarization.\\

At AdA storage ring constructed in LNF, Frascati and brought to LAL,
Orsay (the maximal observed luminosity $L \sim 10^{25}~cm^{-2}s^{-1}$):\\

\hspace{3mm} 1. Discovery of Touschek effect \cite{BCG}.

\hspace{3mm} 2. The first observation of $e^+e^-$ collision (1964)
in the bremsstrahlung reaction $e^+e^-\rightarrow e^+e^- \gamma$ \cite{BCGG}.\\

So, during quite short time new type of accelerator was developed.
This included fast ejection of beam from accelerators which were
used as injectors (where it was necessary), development of channels,
convertors of electron beam into positron one, fast injection of
beams into storage ring, prolong  operation of storage ring with
enough small beam dimensions (to have an acceptable luminosity),
which required high vacuum and damping of many instabilities evolved
during operation.

Both first generation detectors at VEPP-2 and ACO  had some specific
features. 1) A good solid angle. 2) Ability to identify the
particles in an observed event. 3) Reasonable track position
accuracy. 4) Momentum analysis. 5) Background rejection.

The first colliding beam experiments tested QED up to distances more
than 100 times smaller than characteristic hadron dimension.
Described above results completely changed understanding of
electromagnetic structure of hadrons supporting from one side the
basic idea of vector dominance model, but from other side showing
shortages of this model. $SU(3)$ symmetry was tested as well as
$SU(3)$ breaking effects.

Thus, the electron-positron colliding beam project started in INP in
1959 as exotic venture, within a decade  became one of the main
roads of high energy accelerator development. New discoveries were
ahead including November revolution of 1974.

\end{document}